\documentclass[aps,preprint,letterpaper,balancelastpage ,superscriptaddress,showpacs,showkeys,prd]{revtex4}
\usepackage{amsfonts}
\usepackage{amsmath}
\usepackage{amssymb}
\usepackage{eurosym}
\usepackage{paralist}
\usepackage{graphicx}
\usepackage[colorlinks=true,linkcolor=blue,citecolor=blue, filecolor=blue,urlcolor=blue]{hyperref}

\begin{document}

\title{Some cosmological solutions in Einstein-Chern-Simons gravity}

\author{L. Avil\'es}
\email{luisaviles@udec.cl}
\affiliation{Departamento de F\'{\i}sica, Universidad de Concepci\'{o}n, Casilla 160-C,
Concepci\'{o}n, Chile}

\author{P. Mella}
\email{patriciomella@udec.cl}
\affiliation{Departamento de F\'{\i}sica, Universidad de Concepci\'{o}n, Casilla 160-C,
Concepci\'{o}n, Chile}

\author{C. Quinzacara}
\email{cristian.cortesq@uss.cl}
\affiliation{Facultad de Ingenier\'ia y Tecnolog\'ia, Universidad San Sebasti\'an, Lientur 1457, Concepci\'on 4080871, Chile}

\author{P. Salgado}
\email{pasalgad@udec.cl}
\affiliation{Departamento de F\'{\i}sica, Universidad de Concepci\'{o}n, Casilla 160-C,
Concepci\'{o}n, Chile}
\date{\today }

\begin{abstract}
In this paper we find new solutions for the so called Einstein-Chern-Simons
Friedmann-Robertson-Walker field equations studied in refs. \cite{gms,cgqs}. We consider three cases:\begin{inparaenum}[(i)] 
\item in the first case we find some
solutions of the five-dimensional ChS-FRW field equations when the $h^a$ field is a perfect fluid that obeys a barotropic equation of state; 
\item in the second case we study the solutions, for the cases $\gamma =1/2,\ 3/4$, when the $h^a$ field is a five dimensional politropic fluid that obeys the equation $P^{(h)}=\omega ^{(h)}\rho ^{(h)\gamma }$; 
\item in the third case we find the scale factor and the state parameter $\omega (t)$ when the $h^a$ field is a variable modified Chaplygin gas.
\end{inparaenum}

We consider also a space-time metric which contains as a subspace to the usual four-dimensional FRW and then we study the same three cases considered in the five-dimensional, namely when\begin{inparaenum}[(i)] 
\item the $h^a$ field is a perfect fluid, 
\item the $h^a$ field is a five dimensional politropic fluid and
\item the $h^a$ field is a variable modified Chaplygin gas.
\end{inparaenum}
\end{abstract}


\maketitle

\section{Introduction}

The principles underlying the general theory of relativity states that the space-time is a dynamical object which has independent degrees of freedom, and is governed by the Einstein field equations. This means that in General Relativity (GR) the geometry is dynamically determined. Therefore, the construction of a gauge theory of gravity requires an action that does not consider a fixed space-time background. An action for gravity fulfilling these conditions, albeit only in odd-dimensional space-time, $d=2n+1$, was proposed long ago by Chamseddine~\cite{champ1,champ2,zan1}.

If Chern-Simons theories are the appropriate gauge-theories to provide a framework for the gravitational interaction, then these theories must satisfy the correspondence principle, namely they must be related to General Relativity.  

In ref. \cite{salg1} was shown that the standard, five-dimensional General Relativity (without a cosmological constant) can be obtained from Chern-Simons gravity theory for a certain Lie algebra $\mathfrak{B}$. The Chern-Simons Lagrangian is built from a $\mathfrak{B}$-valued, one-form gauge connection $\boldsymbol A$ which depends on a scale parameter $l$ which can be interpreted as a coupling constant that characterizes different regimes within the theory. The $\mathfrak{B}$ algebra, on the other hand, is obtained from the AdS algebra and a particular semigroup $S$ by means of the S-expansion procedure introduced in refs. \cite{salg2,salg3,azcarr}. The field content induced by $\mathfrak{B}$ includes the vielbein $e^{a}$, the spin connection $\omega ^{ab}$ and two extra bosonic fields $h^{a}$ and $k^{ab}$.  

In ref. \cite{salg1} was then shown that it is possible to recover odd-dimensional Einstein gravity theory from a Chern-Simons theory in the limit where the coupling constant $l$ tends to zero while keeping the effective Newton's constant fixed. 

In ref. \cite{gms} was considered a 5-dimensional lagrangian $\mathcal{L=L}_\text{EChS}^{(5)}+\kappa \mathcal{L}_\text{M}$ which is composed of a gravitational sector and a sector of matter, where the gravitational sector is given by the so called Einstein-Chern-Simons gravity action
\begin{equation}
\mathcal{L}_{\mathrm{EChS}}^{\left( 5\right) }=\alpha _{1}l^{2}\epsilon
_{abcde}R^{ab}R^{cd}e^{e}+\alpha _{3}\epsilon _{abcde}\left( \frac{2}{3}%
R^{ab}e^{c}e^{d}e^{e}+2l^{2}k^{ab}R^{cd}T^{e}+l^{2}R^{ab}R^{cd}h^{e}\right)
\label{eins}
\end{equation}
instead of the Einstein-Hilbert action and where the matter sector is given by the so called perfect fluid. In this reference was studied the implications that has on the cosmological evolution, the fact of replacing the Einstein-Hilbert action by the Chern-Simons action in the gravitational sector, for a metric of Friedmann-Robertson-Walker (FRW). Using a compactification procedure known as dynamic compactification, was found that the cosmological field equations obtained from the Chern-Simons gravity theory lead, in a certain limit, to the usual 4-dimensional FRW equations.

It is the purpose of this work to find some new cosmological solutions for the so called Einstein-Chern-Simons-Friedmann-Robertson-Walker field equations, and to show how such new solutions lead, in a certain limit, to the usual cosmological solutions of Einstein theory of gravitation.

To find the new cosmological solutions we interpret the $h^{a}$ field as the dark energy in three different cases:\begin{inparaenum}[(i)]
\item in the first case we interpret the $h^a$ field as a perfect fluid that obeys a barotropic equation of state;
\item in the second case we interpret the $h^a$ field as a
politropic fluid that obeys the equation $P^{(h)}=\omega ^{(h)}\rho^{(h)\gamma }$ and 
\item in the third case we interpret the $h^a$ field as a variable modified Chaplygin gas.
\end{inparaenum}

This paper is organized as follows: In section \ref{sec02} we briefly review the called Einstein-Chern-Simons Friedmann-Robertson-Walker field equations and their solutions when the standard energy-momentum tensor is modeled as a barotropic fluid. In section \ref{sec03} we study three cases where $h^{a}$ field is interpreted as the dark energy and the standard energy-momentum tensor is modeled as a barotropic fluid with variable parameter of state $\omega(t)$. In the first case, we assume that the $h^a$ field is a perfect fluid which obeys the barotropic equation of state $P^{(h)}=\omega ^{(h)}\rho ^{(h)}$, with $\omega$ a constant, and then we find some solutions of the five-dimensional ChS-FRW field equations. In the second case, we study the case when the $h^a$ field is a five dimensional politropic fluid which obeys $P^{(h)}=\omega ^{(h)}\rho^{(h)\gamma}$, with $\omega =\omega(t)$, where some solutions for $\gamma =1/2,\ 3/4$ are found. In the third case, the scale factor and the state parameter $\omega (t)$ are found when the $h^a$ field is a variable modified Chaplygin gas. In section \ref{sec06} a space-time metric which contain as a subspace the usual four-dimensional FRW metric is found and then we studied the same three cases considered in the five-dimensional case. A Summary concludes this work.

\section{Review of five-dimensional EChS-FRW field equations and some of their solutions\label{sec02}}

In this section we review the FRW field equations obtained from the Lagrangian (\ref{eins}) and some of their solutions. From ref. \cite{gms} we know that the EChS-FRW field equations are given by,
\begin{align*}
6\left( \frac{\dot{a}^{2}+k}{a^{2}}\right) +\alpha l^{2}\left(\frac{\dot{a}^{2}+k}{a^{2}}\right) ^{2}&=\kappa _{1}\rho, %\label{80y8}
\\
3\left[ \frac{\ddot{a}}{a}+\left( \frac{\dot{a}^{2}+k}{a^{2}}\right) \right] +\alpha l^{2}\frac{\ddot{a}}{a}\left( \frac{\dot{a}^{2}+k}{a^{2}}\right) &=-\kappa _{1}P, %\label{80y9}
\\
l^{2}\left( \frac{\dot{a}^{2}+k}{a^{2}}\right) ^{2}&=\kappa
_{2}\rho^{(h)}, %\label{noventa}
\\
l^{2}\frac{\ddot{a}}{a}\left( \frac{\dot{a}^{2}+k}{a^{2}}\right) &=-\kappa_{2}P^{(h)},  %\label{90y1}
\\
\left( g-f\right) \frac{\dot{a}}{a}+\dot{g}&=0,
%\label{90y2}
\end{align*}
where $\kappa_1>0$ and $\kappa_2>0$ are appropriate coupling constants and $\alpha=3\alpha _{1}/\alpha _{3}$. Henceforth we assume $\alpha$ is a positive constant. The metric tensor of space-time is given by a FRW metric-type, i.e., $a(t)$ is the cosmic scale factor and $k$ is the curvature of 4-dimensional space. On the other hand, the standard matter is a perfect fluid whose proper energy (or mass) density is $\rho$ and its proper pressure is $P$. Furthermore, $\rho^{(h)}$ and $P^{(h)}$ are the energy density and pressure for the perfect fluid associated to the $h^{a}$ field, also in the comoving frame. The functions $g$ and $f$ are the components of $h^{a}$ field
\begin{equation*}
h^0=f(t)\, e^0,\quad h^i=g(t)\,e^i,\quad i=1,\dots, 4
\end{equation*}
where $e^a$ is the vielbein one-form.

When one consider the case $k=0,$ these equations take the form
\begin{align}
6H^{2}+\alpha l^{2}H^{4}& =\kappa _{1}\rho,  \label{drei} \\
\dot{\rho}+4H(\rho +P)& =0,  \label{vier} \\
l^{2}H^{4}& =\kappa _{2}\rho^{(h)},  \label{funf} \\
\dot{\rho}^{(h)}+4H(\rho^{(h)}+P^{(h)})& =0,  \label{sechs} \\
(g-f)H+\dot{g}& =0,  \label{sieben}
\end{align}%
where we have introduced the Hubble parameter $H=\frac{\dot{a}}{a}$.

\subsection{Standard energy-momentum tensor as a barotropic fluid with constant parameter of state}

Following the same procedure used in general relativity, we further assume that the perfect fluid obeys the barotropic equation of state
\begin{equation}
P=\omega \rho  \label{117}
\end{equation}%
where $\omega$ can be a time-dependent function or a constant. If $\omega $ is a constant then, introducing (\ref{117}) in (\ref{vier}) we obtain
\begin{equation*}
\rho=\rho_{0}\left( \frac{a_{0}}{a}\right) ^{4(\omega+1)}. \label{ciento}
\end{equation*}
The subscript zero means evaluation at the present time $t_{0}=0$.

Using equations (\ref{drei}), (\ref{vier}) and (\ref{117}) we obtain
\begin{equation}
(1+\omega)\left( 6+\alpha l^{2}H^2\right) 
H^2=-\left(3+\alpha l^{2}H^2\right) \dot H.  \label{118}
\end{equation}
This is the equation giving the behavior of the scale factor $a=a(t)$ and is to be solved for the cases when the parameter is uniquely $\omega=-1$ and the general case when $\omega\neq-1$.

\subsubsection{Solutions for $\omega =-1$}

If $\omega=-1$, the equation (\ref{118}) reduces to
\begin{equation*}
\left( 3+\alpha l^{2}\frac{\dot{a}^{2}}{a^{2}}\right) \dot H =0.  \label{119}
\end{equation*}
Since we have assumed that $\alpha >0$ the last equation leads to $H=H_0$, where $H_{0}=H(0)$ is the Hubble constant. This solution leads to
\begin{equation*}
a(t)=a_0e^{H_{0}t}  \label{122}
\end{equation*}%
which is a de Sitter-type solution with $a_0=a(0)$. Since equation (\ref{drei}) is a quadratic equation in $H^2$ its solution yields a value for the Hubble parameter of the form 
\begin{equation}
H^2=-\frac{3}{\alpha l^{2}}\left(1\pm 
\sqrt{ 1+ \frac{\alpha l^{2}}{9}\kappa_{1}\rho }\,\right)
\label{124}
\end{equation}

If we consider the case of small $l^{2}$ limit, we can expand the root to first order in $l^{2}$. In the expansion we can see that it is necessary to take the negative sign in front of square root to recover the FRW equations when $l^{2}=0$ . Thus in first order approximation, equation (\ref{124}) takes the form 
\begin{equation*}
H^{2}=\frac{\dot{a}^{2}}{a^{2}}\approx \frac{\kappa_{1}\rho }{6}
\left(1-\frac{\alpha l^{2}}{36}\,\kappa_{1}\rho \right) \label{125}
\end{equation*}%
or 
\begin{equation*}
H_{0}^{2}\approx \frac{\kappa _{1}\rho _{0}}{6}\left(1-\frac{\alpha l^{2}}{36}\,\kappa _{1}\rho _{0}\right) \label{126}
\end{equation*}%
expression that in the $l^{2}=0$ limit is identical to that obtained when using the Einstein equations in five dimensions.

\subsubsection{Solutions for $\omega \neq -1$}

We consider now the behavior of the scale factor for the general case when $\omega$ is left as a free parameter. By integrating equation (\ref{118}) with $\omega \neq -1$ we obtain
\begin{equation}
\frac{\dot{a}}{a}=\frac{1}{\gamma}\tan \left\{ \frac{1}{\gamma }%
\biggl( \frac{a}{\dot{a}}-2(1+\omega )\left(t-t'\right)\biggr) \right\}
\label{123}
\end{equation}%
where
\begin{equation*}
t'=\frac{1}{2H_0(1+\omega)}\Bigl(\gamma H_0\arctan(\gamma H_0)-1\Bigr)
\end{equation*}
and we have defined $\gamma =\sqrt{\alpha l^{2}/6}$. In the small $l^{2}$ limit, we can expand equation (\ref{123}). In fact, taking the $\arctan$ of each side of (\ref{123}), carrying out the expansion to first order in $\gamma$ and solving for the Hubble parameter, we obtain
\begin{equation}
\frac{\dot{a}}{a}\approx -\frac{(1+\omega )}{\gamma ^{2}}\left(t-t'\right)\left(
1\pm \sqrt{1+\left( \frac{\gamma }{(1+\omega )\left(t-t'\right)}\right) ^{2}}\right).
\label{129}
\end{equation}
Expanding the square root in (\ref{129}) to second order in $\gamma^2=\alpha l^{2}/6$, considering the negative sign to recover the five-dimensional FRW equations, and integrating, we obtain a value for the scale factor of the form
\begin{equation}
a(t)\approx C(t-t')^{1/2(1+\omega)}\left(1+\frac{\alpha l^{2}}{12}\left( \frac{1}{2(1+\omega )}\right) ^{3}\left(\frac{1}{(t-t')^{2}}-\frac{1}{{t'}^{2}}\right)\right)  \label{131}
\end{equation}%
where
\begin{equation*}
C=a_0\,{\vert t'\vert}^{-1/2(1+\omega)}\approx a_0\Bigl(2H_0(1+\omega)\Bigr)^{1/2(1+\omega)}\approx a_{0}\left( \frac{2}{3}\kappa_{1}\rho_{0}\,(1+\omega ) ^{2}\right)^{1/4(1+\omega )}.
\end{equation*}

From equation (\ref{131}) we see that:\begin{inparaenum}[(i)] \item The cases of greatest physical interest are those with $\omega =0$ and $\omega =1/4$, which are in the category $\omega \neq -1$. These cases are usually called the eras of matter and radiation respectively.
\item For small values of $l^{2}$ and for values not small of $t^{2}$ we have that the term the right in (\ref{131}) is negligible compared to the first and we recover the usual solutions to the 5-dimensional FRW equations.
\item In case that $t^{2}$ is the order of $l^{2}$ we have that the term on the right in (\ref{131}) is not negligible compared to the first and therefore becomes important in the description of evolution: this is a notable difference with the results obtained from general relativity. If the term on the right in (\ref{131}) takes a value greater than zero, then it is possible that this term is important for the description of an inflationary period of the universe.
\item We should note that this solution corresponds to a valid theory in five dimensions which describes the evolution of 5-dimensional space-time.
\end{inparaenum}

\section{Five-dimensional EChS-FRW field equations: Standard ener\-gy-mo\-men\-tum tensor as a barotropic fluid with variable parameter of state\label{sec03}}

In this section we will assume that the standard matter perfect fluid is a barotropic fluid with variable parameter of state, i.e., which obeys the equation of state 
\begin{equation}
P=\omega (t)\rho.  \label{eosm}
\end{equation}%
Here $\omega $ can be a time-dependent function. We will study the behavior of the scale factor $a=a(t)$, the energy density and the variable parameter $\omega(t)$.

\subsection{The $h^a$ fluid as a barotropic fluid with constant parameter of state}

Now, we assume that the perfect fluid asociated to $h^{a}$ field obeys the barotropic equation of state 
\begin{equation}
P^{(h)}=\omega ^{(h)}\rho ^{(h)},  \label{eosh}
\end{equation}%
where $\omega ^{(h)}$ is a constant parameter of state. We will solve for the cases when the parameter is uniquely $\omega ^{(h)}=-1$ and the general case when $\omega ^{(h)}\neq -1$.

\subsubsection{Solutions for $\omega^{(h)}=-1$}

If $\omega ^{(h)}=-1$, the equation (\ref{sechs}) with the condition (\ref{eosh}) reduces to 
\begin{equation}
\dot{\rho}^{(h)}=0.  \label{seven}
\end{equation}%
From (\ref{funf}) and (\ref{seven}) we find $H(t)=H_{0}$ where $H_0=H(0)$, whose solution is given by
\begin{equation}
a(t)=a_{0}e^{H_{0}t}  \label{sfw1}
\end{equation}%
where $a_{0}=a(0)$, which is a de Sitter-type solution.

Introducing (\ref{sfw1}) or $H(t)=H_{0}$ in equations (\ref{funf}) and (\ref{drei}) we find
\begin{equation*}
\rho^{(h)}(t)=\frac{l^{2} H_{0}^{4}}{\kappa_{2}},\quad
\rho(t)=H_{0}^{2}\left(\frac{6+l^{2}\alpha H_{0}^{2}}{\kappa_{1}}\right).
\label{rhow1}
\end{equation*}

Finally, from (\ref{eosm}) and (\ref{vier}) we can see that $\omega(t) =-1$. Therefore, if $\omega ^{(h)}=-1$ then there is not a $\omega (t)$ variable. Hence both are constant. It is interesting to note that in the $l^{2}=0$ limit, we get the same result that one can obtain using the Einstein equations in five dimensions.

\subsubsection{Solutions for $\omega^{(h)}\neq -1$}

We consider now the behavior of the scale factor for the case when $\omega^{(h)}\neq -1$.  If $\omega ^{(h)}\neq -1$, the equation (\ref{sechs}) with the condition (\ref{eosh}) reduces to
\begin{equation}
\dot{\rho}^{(h)}+4H\rho ^{(h)}(\omega ^{(h)}+1)=0.  \label{elf}
\end{equation}
From (\ref{funf}) and (\ref{elf}) we find
\begin{equation}
H(t)=\frac{H_{0}}{1+ \left( \omega^{(h)}+1\right)H_{0}t
},  \label{hubble5}
\end{equation}
where $H_{0}=H(0)$. From (\ref{hubble5}) we find that 
\begin{equation}
a(t)=a_{0}\left( 1+(\omega ^{(h)}+1)H_{0}t\right) {}^{\frac{1}{\omega
^{(h)}+1}}  \label{sf5}
\end{equation}
where $a_{0}=a(0)$. Introducing (\ref{sf5}) or (\ref{hubble5}) in equations (\ref{funf}) and (\ref{drei}) we have
\begin{align*}
\rho^{(h)}(t)&=\frac{l^{2}H_{0}^{4}}{\kappa_{2}\left( 1+(\omega^{(h)}+1)H_{0}t%
\right)^{4}},  %\label{rhoh5}
\\
\rho(t)&=\frac{H_{0}^{2}\left( 6\Bigl(1+ \left( \omega^{(h)}+1\right)H_{0}t
\Bigr)^{2}+\alpha l^{2}H_{0}^{2}\right) }{k_{1}\Bigl(1+ \left( \omega^{(h)}+1\right)H_{0}t
\Bigr)^{4}}.
\end{align*}
Finally, from (\ref{eosm}) and (\ref{vier}) we can see that
\begin{equation*}
\omega (t)=\frac{\alpha H_{0}^{2}l^{2}\omega ^{(h)}+3(\omega
^{(h)}-1)\Bigl(1+ \left( \omega^{(h)}+1\right)H_{0}t
\Bigr)^{2}}{\alpha
H_{0}^{2}l^{2}+6\Bigl(1+ \left( \omega^{(h)}+1\right)H_{0}t
\Bigr)^{2}}.
\end{equation*}

It should be mentioned that if $\omega^{(h)}<1$, then the equation (\ref{sf5}) has a similar behavior to the equation which describes the acceleration of the universe due to the presence of the phantom energy, and has a big rip-like future singularity at time
\begin{equation*}
t_{r}=-\frac{1}{(\omega^{(h)}+1)H_{0}}.
\end{equation*}
Note that in the $l^{2}=0$ limit, we get the results of the five-dimensional standard cosmology for a scenario type big rip where:\begin{inparaenum}[(i)]
\item the state parameter $\omega (t)=(\omega ^{(h)}+1)/2$ is constant and
\item the scale factor is given as an ansatz.
\end{inparaenum}

\begin{figure}[t]%figure1
\centering
\includegraphics[width=0.8\columnwidth]{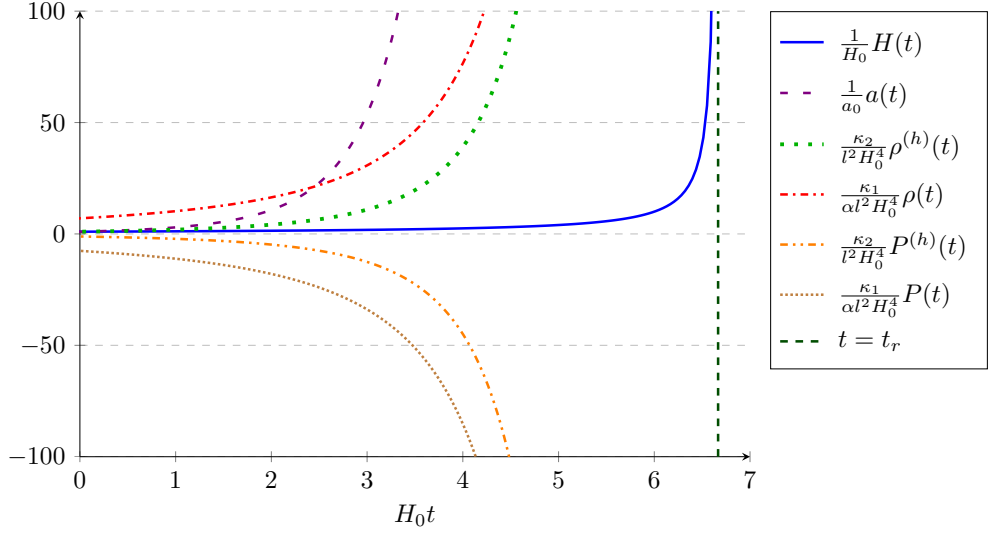}
\caption{Behavior of different cosmological parameters (vertical axis) in terms of $t$, with $\protect{\omega^{(h)}=-1.15}$, and setting $H_0$, $l$ and $\alpha$ to $1$. Here we have a big rip-like singularity at time $H_0t_r=20/3$. \label{fig01}}
\end{figure}

\begin{figure}%figure2
\centering
\includegraphics[width=0.8\columnwidth]{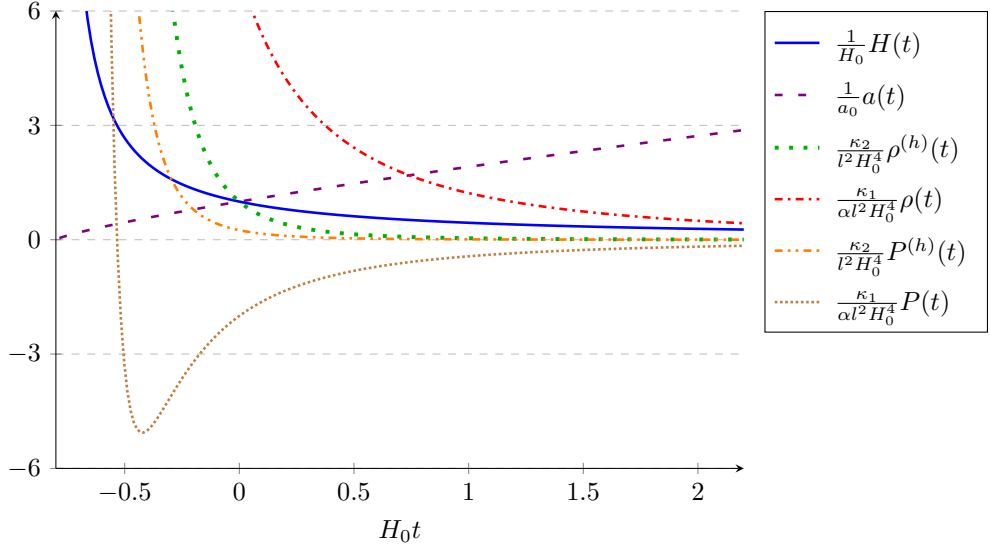}
\caption{Behavior of different cosmological parameters (vertical axis) in terms of $t$, with $\omega^{(h)}=1/4$, and setting $H_0$, $l$ and $\alpha$ to $1$. Here we have $a=0$ at time $H_0t_b=-1/(\omega^{(h)}+1)=-4/5$. \label{fig02}}
\end{figure}

In figure \ref{fig01} we can see the behavior of the cosmological parameters in term of time, $\omega ^{(h)}=-1.15$, with $H_{0}t_{r}=20/3$. When $-1<\omega^{(h)}<-1/2$ we have not big rip but an accelerated expansion (i.e., $\ddot{a}(t)>0$) that tends to a constant in the future. For $\omega^{(h)}=-1/2$ we have a constant accelerated expansion, i.e., $\ddot{a}(t)=\text{cte}$. For $-1/2<\omega ^{(h)}<0$ we have a decelerated expansion, i.e., ($\ddot{a}(t)<0 $). If $\omega^{(h)}=0$ then we have a solution without accelerated expansion (i.e., $\ddot{a}(t)=0$). For $0<\omega ^{(h)}<1$ again we have a decelerated expansion. In figure \ref{fig02} we can see the behavior of the cosmological parameters in term of the time, for intance, radiation $\omega^{(h)}=1/4$.

\begin{figure}%figure3
\centering
\includegraphics[width=0.7\columnwidth]{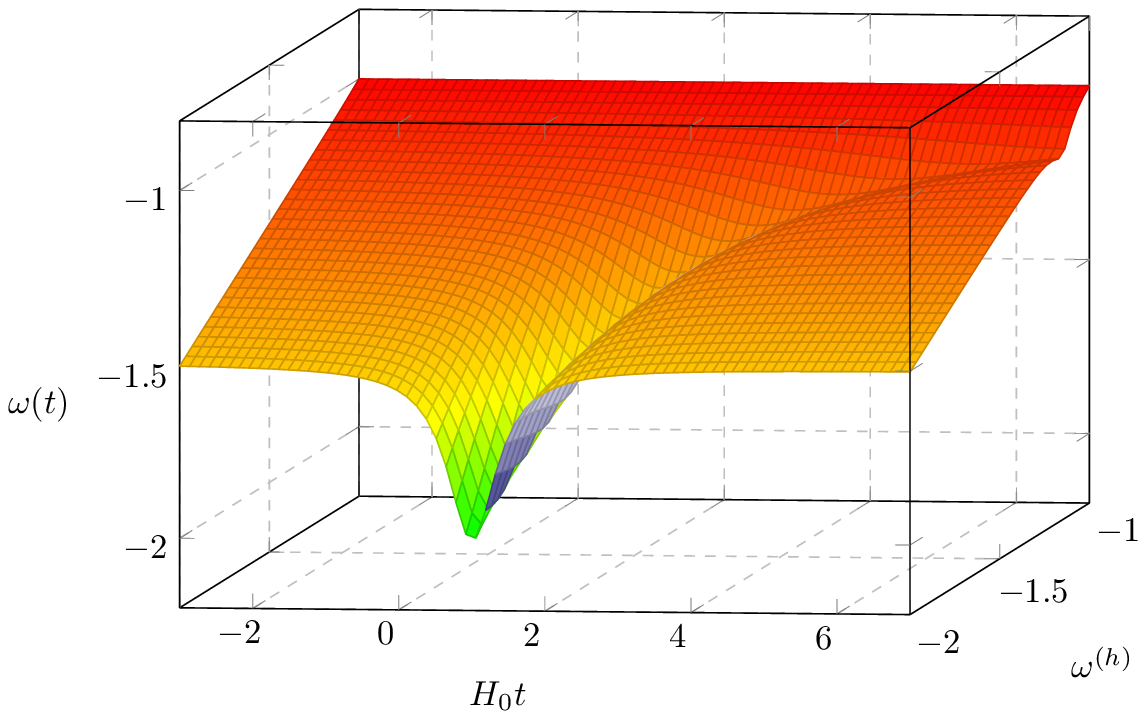}
\caption{Behavior of $\omega(t)$ in terms of $t$ and $-2<\omega^{(h)}<-1$, and setting $H_0$, $l$ and $\alpha$ to $1$. \label{fig03}}
\end{figure}

\begin{figure}%figure4
\centering
\includegraphics[width=0.7\columnwidth]{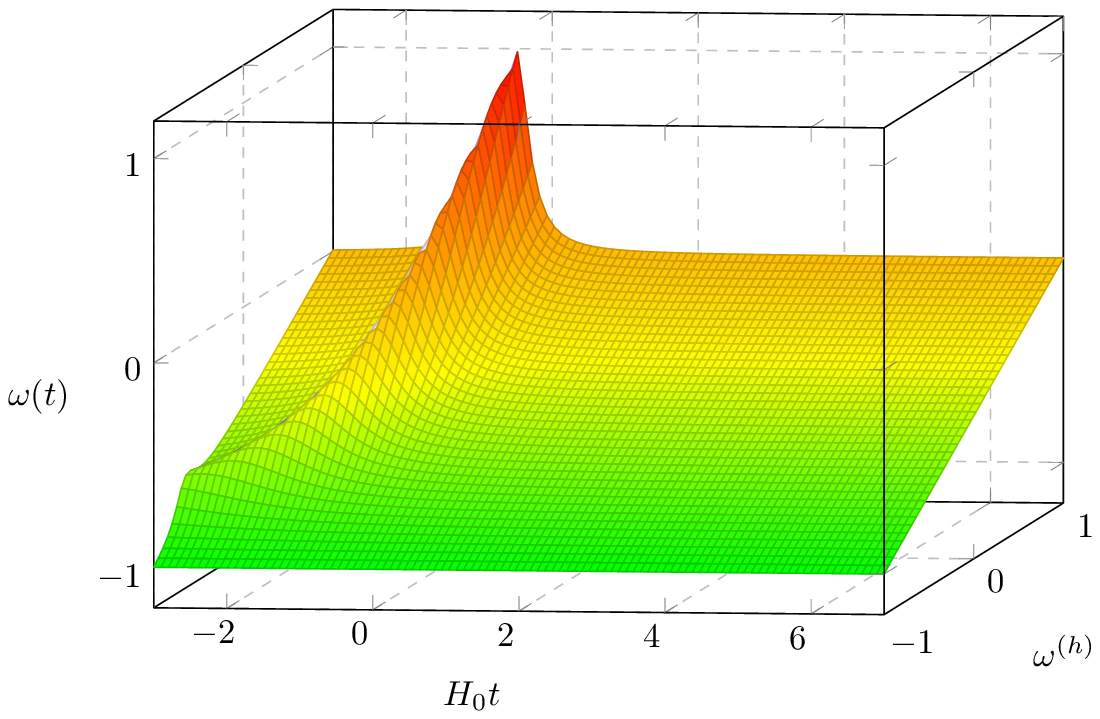}
\caption{Behavior of $\omega(t)$ in terms of $t$ and $-1<\omega^{(h)}<1$, and setting $H_0$, $l$ and $\alpha$ to $1$.\label{fig04}}
\end{figure}

Finally in figure \ref{fig03} and figure \ref{fig04} we show the behavior of $\omega (t)$ in terms of time and $\omega ^{(h)}$. Here we can see that for any value of phantom-like parameter of state, i.e., $\omega ^{(h)}<-1$, we have a phantom-like variable parameter of state $\omega (t)$. When $-1<\omega^{(h)}<0.75$\ , the variable parameter of state $\omega (t)<0$, and when $0.75<\omega ^{(h)}<1$, the variable parameter of state $\omega (t)$ evolves quickly from positive to negative values.

It might be interesting to mention that in this case we have obtained the exact form of the scale factor without having to make an expansion to the scale factor as was done in ref. \cite{gms}. This means that we have found a different solution that complements that found for $\omega \neq -1$ in ref. \cite{gms}.

\subsection{The $h^a$ fluid as a politropic fluid\label{sec04}}

Now we consider the case where the field $h^{a}$ can be a polytropic fluid modelling dark energy. The corresponding state equation is given by 
\begin{equation}
P^{(h)}=\omega ^{(h)}\rho ^{(h)\gamma },  \label{eosp}
\end{equation}%
where, in analogy to four-dimensional standard cosmology, the state parameter $\omega ^{(h)}$ can take constant values and $\gamma$ is the polytropic index with $\gamma >0$ and $\gamma \neq 1$ \cite{poly1,poly2,poly3,poly4,poly5,poly6}. When $\gamma =1$, the equation of state is reduced to the barotropic equation of state. The matter field will be modeled as a fluid whose state equation with variable parameter, is given by equation (\ref{eosm}). In this subsection we study two cases having exact solution: $\gamma =\frac{1}{2}$ and $\gamma =\frac{3}{4}$.

\subsubsection{Case 1: $\gamma =\frac{1}{2}$}

From equations (\ref{funf}), (\ref{sechs}) and (\ref{eosp}) we found 
\begin{equation}
\dot{H}+H^{2}+\omega ^{(h)}\left( \frac{\kappa _{2}}{l^{2}}\right)^{\frac{1}{2}}=0,  \label{poli1}
\end{equation}%
where we can see that it is not possible to consider the limit $l=0$. To find a solution for (\ref{poli1}), we have three different cases. First, when $\omega ^{(h)}>0$ the scale factor have a sector where is negative, so that we discard this solution. Second, if $\omega^{(h)}=0$ the scale factor is linear in $t$, so that the solution is a non-accelerated expansion. When $\omega ^{(h)}<0$, we find a solution given by 
\begin{equation}
H(t)=H_1\tanh \Bigl( H_1\left( t-t_1\right) \Bigr)  \label{poli2}
\end{equation}
where the constants $H_1$ and $t_1$ are given by
\begin{equation*}
H_1=\sqrt{-\omega ^{(h)}}\left( \frac{\kappa_{2}}{l^{2}}\right) ^{\frac{1}{4}},\quad t_1=-\frac{1}{H_1}\,\text{arctanh}\left(\frac{H_0}{H_1}\right).
\end{equation*}

From equation (\ref{poli2}) it is direct to see that
\begin{equation*}
a(t)=\frac{a_0}{\cosh(H_1t_1)}\cosh \Bigl( H_1\left( t-t_1\right) \Bigr).  \label{poli3}
\end{equation*}

Introducing (\ref{poli2}) in equations (\ref{funf}) and (\ref{drei}) we have%
\begin{align*}
\rho ^{(h)}(t)&=\left(\omega ^{(h)}\right)^{2}\tanh^{4}\Bigl( H_1\left( t-t_1\right) \Bigr),  %\label{poli4}
\\
\rho (t)&=\frac{-\omega ^{(h)}}{\kappa_{1}}\tanh ^{2}\Bigl( H_1\left( t-t_1\right) \Bigr) \left(6\sqrt{\frac{\text{$\kappa $}_{2}}{l^{2}}}- \alpha\kappa_{2}\, \omega ^{(h)} \tanh^{2}\Bigl( H_1\left( t-t_1\right) \Bigr)\right).
%\label{poli5}
\end{align*}

Finally, from (\ref{eosm}) and (\ref{vier}) we can see that%
\begin{equation*}
\omega (t)=-\frac{3\sqrt{\frac{\kappa_{2}}{l^{2}}}\text{csch}^{2}\Bigl( H_1\left(t-t_1\right) \Bigr) \cosh \Bigl(2 H_1\left( t-t_1\right) \Bigr)-\alpha \kappa_{2}\,\omega
^{(h)}}{6\sqrt{\frac{\text{$\kappa $}_{2}}{l^{2}}}- \alpha\kappa_{2}\, \omega ^{(h)} \tanh^{2}\Bigl( H_1\left( t-t_1\right) \Bigr)}.  \label{poli6}
\end{equation*}

\subsubsection{Caso 2: $\gamma =\frac{3}{4}$}

From equations (\ref{funf}), (\ref{sechs}) and (\ref{eosp}) we found the following equation 
\begin{equation}
\dot{H}+H^{2}+\omega ^{(h)}\left( \frac{\kappa _{2}}{l^{2}}\right) ^{\frac{1%
}{4}}H=0,  \label{poli7}
\end{equation}%
where we can see that it is not possible to consider the limit $l=0 $. The solution is given by 
\begin{equation}
H(t)=\frac{H_2}{e^{H_2(t-t_2)}-1}  \label{poli8}
\end{equation}%
with
\begin{equation*}
H_2=\omega ^{(h)}\left( \frac{\kappa _{2}}{l^{2}}\right) ^{\frac{1}{4}},\quad t_2=-\frac{1}{H_2}\ln\left(1+\frac{H_2}{H_0}\right).
\end{equation*}
This means that the scale factor is given by
\begin{equation}
a(t)=a_0\left(\frac{1-e^{-H_2(t-t_2)}}{1-e^{H_2t_2}}\right). \label{fepoli34}
\end{equation}

Note that in order to have a scale factor that grows with time, $\omega^{(h)}$ must be less than zero so that $H_2<0$.

From (\ref{funf}) and (\ref{fepoli34}) or (\ref{poli8}) one find that $\rho^{(h)}$ takes the form 
\begin{equation*}
\rho^{(h)}(t)=\left( \frac{\omega^{(h)}}{e^{H_2(t-t_2)}-1}\right) ^{4}
\label{poli9}
\end{equation*}
and from equation (\ref{drei}) we can see that
\begin{equation*}
\rho(t)=\frac{\left(6\sqrt{\frac{\kappa_{2}}{l^{2}}}\left(e^{H_2(t-t_2)}-1\right) ^{2}+\alpha\kappa_{2}\left(\omega^{(h)}\right)^2\right)\left(\omega^{(h)}\right)^2}{\kappa_{1}\left(e^{H_2(t-t_2)}-1\right)^{4}} ,  \label{poli10}
\end{equation*}
so that the state parameter $\omega(t)$ is given by 
\begin{equation*}
\omega(t)=\frac{\left( 3\sqrt{\frac{\kappa_{2}}{l^{2}}}\bigl(e^{H_2(t-t_2)}-1\bigr)\bigl(e^{H_2(t-t_2)}-2\bigr) +\alpha\kappa_{2}\left(\omega^{(h)}\right)^2\right) \left(e^{H_2(t-t_2)}-1\right) }{6\sqrt{\frac{\kappa_{2}}{l^{2}}}\left(e^{H_2(t-t_2)}-1\right)^{2}+\alpha\kappa_{2}\left(\omega^{(h)}\right)^2}
\label{poli11}
\end{equation*}

Note that when $\omega^{(h)}=0$ in (\ref{poli7}), we have that the scale factor is linear in $t$, so that the corresponding solution is a expansion non-accelerated.

\subsection{The $h^a$ fluid as a variable modified Chaplygin gas\label{sec05}}

Now consider the case where the field $h^{a}$ can be a variable modified Chaplygin Gas \cite{chap1,chap2,chap3,chap4,chap5,chap6,chap7,chap8,chap9,chapkk} modelling dark energy, in an analogous manner as it occurs in 4-dimensional standard cosmology. The corresponding state equation is given by 
\begin{equation}
P^{(h)}=-\frac{a(t)^{-4\gamma -3}}{\rho ^{(h)\gamma }}  \label{eosgcgm}
\end{equation}
where $0<\gamma <1$. The matter field will be modeled as a fluid whose state equation with variable parameter, is given by equation (\ref{eosm}).

From equations (\ref{sechs}) and (\ref{eosgcgm}) it is direct to see that 
\begin{equation*}
\dot{z}+4\frac{\dot{a}}{a}(\gamma +1)z=4(\gamma +1)\frac{\dot{a}}{%
a^{4(\gamma +1)}},
\end{equation*}
where $z=\rho^{(h)\gamma +1}$. This means that $\rho ^{(h)}$ is given by 
\begin{equation*}
\rho ^{(h)}(t)=\left( a_3a^{-4(\gamma+1)}+4(\gamma+1)a^{-(4\gamma +3)}
\right)^{\frac{1}{\gamma +1}}  \label{dreizehen}
\end{equation*}
here $a_3$ is a constant of integration to be determined from initial conditions.

The equations (\ref{funf}) and (\ref{sechs}) allow us to find the following scale factor 
\begin{equation*}
a(t)=\frac{A(t)^{4(\gamma+1)}-a_3}{4(\gamma +1)}
\end{equation*}
where
\begin{equation*}
A(t)=\bigl((4\gamma+3)H_3(t-t_3)\bigr)^{\frac{1}{4\gamma+3}},\quad H_3=\left(\frac{\kappa_2}{l^2}\right)^{\frac{1}{4}},\quad
t_3=-\frac{\bigl(a_3+4(\gamma+1)a_0\bigr)^{\frac{4\gamma+3}{4(\gamma+1)}}}{(4\gamma+3)H_3}.
\end{equation*}
This result leads to obtaining the following Hubble parameter: 
\begin{equation*}
H(t)=\frac{4(\gamma+1)H_3A(t)}{A(t)^{4(\gamma +1)}-a_{3}}
\end{equation*}%
where we determine the constant $a_3$
\begin{equation*}
a_3=\left(a_0\,\frac{H_0}{H_3}\right)^{4(\gamma+1)}-4(\gamma+1)a_0.
\end{equation*}

On the other hand, using equation (\ref{drei}) is straightforward to obtain 
\begin{equation*}
\rho (t)=\frac{256\alpha l^2(\gamma +1)^{4}H_3^4A(t)^{4}
}{\kappa _{1}\left( A(t)^{4(\gamma +1)}-a_3\right)^{4}}+\frac{96(\gamma +1)^{2}H_3^2A(t)^{2}}{\kappa _{1}\left( A(t)^{4(\gamma +1)}-a_3\right)^{2}}.
\end{equation*}%
This result leads to the following state parameter $\omega (t)$
\begin{equation*}
\omega(t)=\frac{\Bigl((4\gamma+3)A(t)^{4(\gamma+1)}+ a_3\Bigr)\Bigl(3\left(A(t)^{4(\gamma+1)}-a_3\right)^2+16\alpha l^2H_3^2(\gamma+1)^2A(t)^2\Bigr)}{8(\gamma+1)A(t)^{4(\gamma+1)}\Bigl(3\left(A(t)^{4(\gamma+1)}-a_3\right)^2+8\alpha l^2H_3^2(\gamma+1)^2A(t)^2\Bigr)}-1.
\end{equation*}%

\section{Four-dimensional EChS-FRW field equations: Stan\-dard ener\-gy-mo\-men\-tum tensor as a barotropic fluid with variable parameter of state \label{sec06}}

So far we have found some solutions for flat cosmological field equations, which were obtained from a Lagrangian for a Chern-Simons gravity theory, studied in ref. \cite{salg1}. One problem with these solutions is that they are valid only in a five-dimensional space. Now we consider a space-time metric which contains as a subspace the usual FRW metric in four dimensions. Following refs. \cite{andrew,mohamedi} we consider the following five-dimensional metric:

\begin{equation*}
ds^{2}=-dt^{2}+a^{2}(t)\Bigl((dx_{1})^{2}+(dx_{2})^{2}+(dx_{3})^{2}\Bigr)+b^{2}(t)dx^{2}.
\label{frw54}
\end{equation*}

Using the compactification procedure developed in ref. \cite{mohamedi} we have that the scale factor $b(t)$ is given by
\begin{equation*}
b(t)=\frac{1}{a(t)^{n}},\quad n>0,  \label{red}
\end{equation*}%
where the parameter $n$ must be positive for dynamical compactification to take place. Therefore, $b(t)$ gets smaller as the radius of our universe $a(t)$ become bigger. So that the corresponding field equations are given by \cite{gms}

\begin{align}
3(1-n)H^{2}+12\epsilon nH^{4}& =\frac{\kappa _{1}\rho }{2}, \label{fe5da41}
\\
\dot{\rho}+3H(\rho +\widetilde{P})& =0, \label{fe5da42} \\
\epsilon nH^{4}& =\kappa _{2}\rho ^{(h)}, \label{fe5da4} \\
\dot{\rho}^{(h)}+3H(\rho ^{(h)}+\widetilde{P}^{(h)})& =0,  \label{fe5da45} \\
\dot{q}-\frac{n}{3}H(q-f)& =0,  \label{fe5da47} \\
\dot{g}+H(g-f)& =0 \label{fe5da48}
\end{align}
where $\kappa_1>0$ and $\kappa_2>0$ are coupling constants and $\epsilon=\frac{\alpha _{1}}{4\alpha_{3}}l^2$. Henceforth we assume $\epsilon>0$. $\widetilde P$ and $\widetilde P^{(h)}$ are the effective pressures defined in ref. \cite{gms}. The functions $f(t)$, $g(t)$ and $q(t)$ are the components of 5-dimensional $h^a$ field
\begin{equation*}
h^0=f(t)\, e^0,\quad h^i=g(t)\,e^i,\quad h^4=q(t)\,e^4,\quad i=1,\dots,3
\end{equation*}
where $e^a$ is the 5-dimensional vielbein one-form.

\subsection{The $h^a$ fluid as a barotropic fluid with constant parameter of state}

Following the same procedure used in the 5D case, we assume that the dark energy fluid $h^{a}$ obeys the barotropic equation of state
\begin{equation}
\widetilde{P}^{(h)}=\widetilde{\omega}^{(h)}\rho^{(h)}  \label{eosh5d4}
\end{equation}
where the state parameter $\widetilde{\omega}^{(h)}$ is a constant. The matter perfect fluid obeys the equation of state 
\begin{equation}
\widetilde{P}=\widetilde{\omega}(t)\rho.  \label{eosm5d4}
\end{equation}
where the state parameter $\widetilde{\omega}(t)$ can be a time-dependent function.

In this section we study the behavior of the scale factor $a=a(t)$, the energy density and the variable state parameter. We will solve for the cases when the parameter is $\widetilde{\omega}^{(h)}=-1$ and when $\widetilde{\omega}^{(h)}\neq-1$.

\subsubsection{Solutions for $\widetilde{\omega }^{(h)}=-1$ case}

Writing equation (\ref{fe5da45}) in terms of the parameter of Hubble (\ref{fe5da4}) and using the equation of state (\ref{eosh5d4}), we find
\begin{equation*}
H(t)=H_{0}  \label{hh5d4}
\end{equation*}
where $H_{0}=H(0)$, whose de Sitter-type solution is given by
\begin{equation}
a(t)=a_{0}e^{tH_{0}}  \label{ssf5d4}
\end{equation}
with $a_{0}=a(0)$. Introducing (\ref{ssf5d4}) in (\ref{fe5da4})
and (\ref{fe5da41}), we find
\begin{align*}
\rho^{(h)}(t)&=\frac{H_{0}^{4}n\epsilon}{\kappa_{2}},  %\label{rrhoh5d4}
\\
\rho(t)&=\frac{6H_{0}^{2}}{\kappa_{1}}\Bigl( n\left( 4H_{0}^{2}\epsilon-1\right) +1%
\Bigr) .  
%\label{rrho5d4}
\end{align*}

Finally, from (\ref{eosm5d4})$\ $\ and (\ref{fe5da42}) we can see that
\begin{equation*}
\widetilde{\omega}(t)=-1,  \label{ssp5d4}
\end{equation*}
This means that the state parameter $\widetilde{\omega}(t)$ is not a time-dependent function. So that the state parameters $\widetilde{\omega}(t)$ and $\widetilde{\omega}^{(h)}$ are constants. We must add here that this result is consistent with that found in ref. \cite{gms} for $\omega = -1$.

\subsubsection{Solutions for $\widetilde{\omega }^{(h)}\neq-1$ case}

Writing equation (\ref{fe5da45}) in terms of the parameter of Hubble and using the equation of state (\ref{eosh5d4}), we find
\begin{equation}
H(t)=\frac{H_{0}}{1+\frac{3}{4}\left(\widetilde{\omega}^{(h)}+1\right) H_{0}t}
\label{h5d4}
\end{equation}
where $H_{0}=H(0)$. From (\ref{h5d4}) we find that
\begin{equation}
a(t)=a_{0}\left(1+\frac{3}{4}\left( \widetilde{\omega}^{(h)}+1\right) H_{0}t\right)^{\frac{4}{3\left( \widetilde{\omega}^{(h)}+1\right) }},  \label{sf5d4}
\end{equation}
whith $a_{0}=a(0)$. Introducing (\ref{sf5d4}) or (\ref{h5d4}) in (\ref{fe5da4}) and (\ref{fe5da41}), we have

\begin{align*}
\rho^{(h)}(t)&=\frac{\epsilon n H_{0}^{4}}{\kappa_{2}\Bigl( 1+\frac{3}{4}\left( 
\widetilde{\omega}^{(h)}+1\right) H_{0}t\Bigr)^{4}}, %\label{rhoh5d4}
\\
\rho(t)&=\frac{6\, H_{0}^{2}\left( (1-n)\Bigl(1+\frac{3}{4}\left( 
\widetilde{\omega}^{(h)}+1\right)H_{0}t\Bigr) ^{2}+4H_{0}^{2}\epsilon n\right) 
}{\text{$\kappa$}_{1}\Bigl(1+ \frac{3}{4}\left( \widetilde{\omega}^{(h)}+1\right)H_{0}t \Bigr)^{4}}.
%\label{rho5d4}
\end{align*}

Finally, from (\ref{eosm5d4}) and (\ref{fe5da42}) we can see that
\begin{equation*}
\widetilde{\omega}(t)=\frac{(1-n)(\widetilde{\omega}^{(h)}-1)\Bigl(1+\frac{3}{4}(\widetilde{\omega}^{(h)}+1)H_{0}t\Bigr)^{2}+8H_{0}^{2}\epsilon n\,\widetilde{\omega}^{(h)}}{2\left( (1-n)\Bigl(1+\frac{3}{4}(\widetilde{\omega}^{(h)}+1)H_{0}t\Bigr)^{2}+4H_{0}^{2}\epsilon n\right) }.  \label{sp5d4}
\end{equation*}

From equations (\ref{h5d4}) and (\ref{sf5d4}) we can see that:
\begin{enumerate}[(i)]
\item For $0<n<1$, for example $n=1/2,$ we have that in the case $\widetilde{\omega}^{(h)}<-1$, the equation (\ref{sf5d4}) has a similar behavior to the equation which describes an accelerated universe with a big rip-like future singularities at time
\begin{equation*} %\label{tr01}
t_{r}=-\frac{4}{3(\widetilde{\omega }^{(h)}+1)H_{0}}.
\end{equation*}
See figures \ref{fig05} and \ref{fig06} for the behavior of the cosmological parameters with $\widetilde{\omega }^{(h)}=-1.15$ and $H_{0}t_{r}=80/9$.

\begin{figure}[tbp]
%figure5
\centering
\includegraphics[width=0.8\columnwidth]{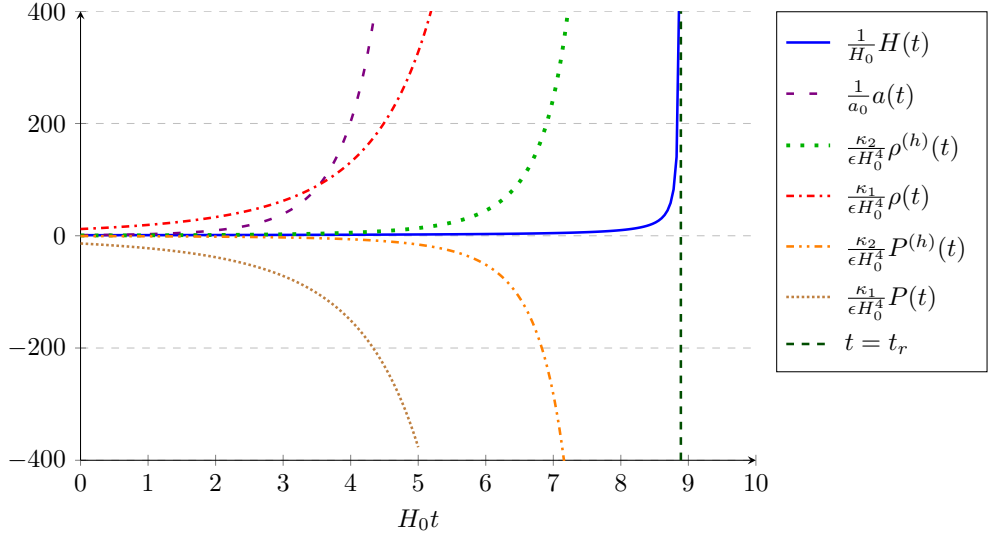}
\caption{Behavior of different cosmological parameters (vertical axis) in
terms of $t$, with $\protect{\widetilde{\omega}^{(h)}=-1.15}$, $n=1/2$, $\epsilon=1$ and $H_0=72$ \cite{hub1,hub2,hub3,hub4}. Here we have a big rip-like singularity at time $H_0t_r=80/9$.\label{fig05}}
\end{figure}

\begin{figure}[tbp]
%figure6
\centering
\includegraphics[width=0.7\columnwidth]{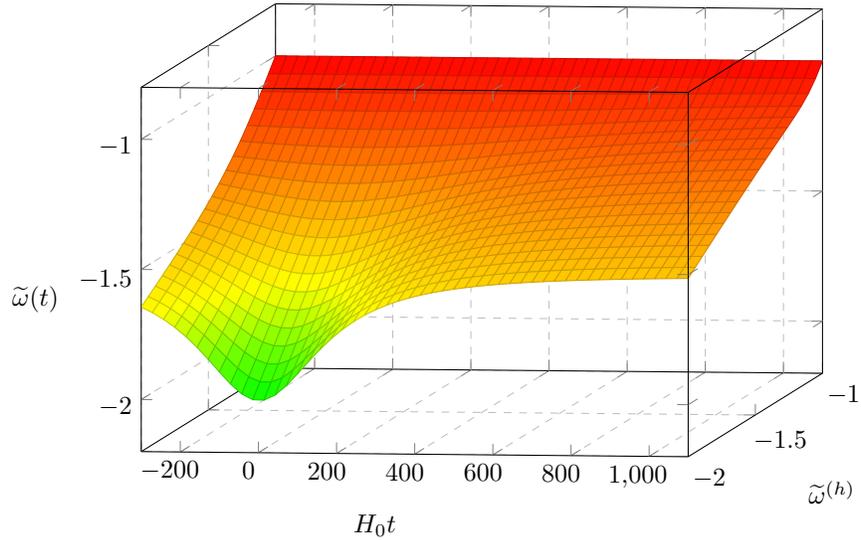}
\caption{Behavior of $\widetilde{\omega}(t)$ in terms of $t$
and $-2<\widetilde{\omega}^{(h)}<-1$, with $n=1/2$, $\epsilon=1$ and $H_0=72$.\label{fig06}}
\end{figure}

Now when $-1<\widetilde{\omega}^{(h)}<0$, there is not big rip. See 
figures \ref{fig07} and \ref{fig08} for an example with $\widetilde\omega^{(h)}=1/3$. 

\item For $n=1$ we have $\widetilde{\omega}(t)=\widetilde{\omega}%
^{(h)}=\text{constant}$.

\begin{figure}[tbp]
%figure7
\centering
\includegraphics[width=0.8\columnwidth]{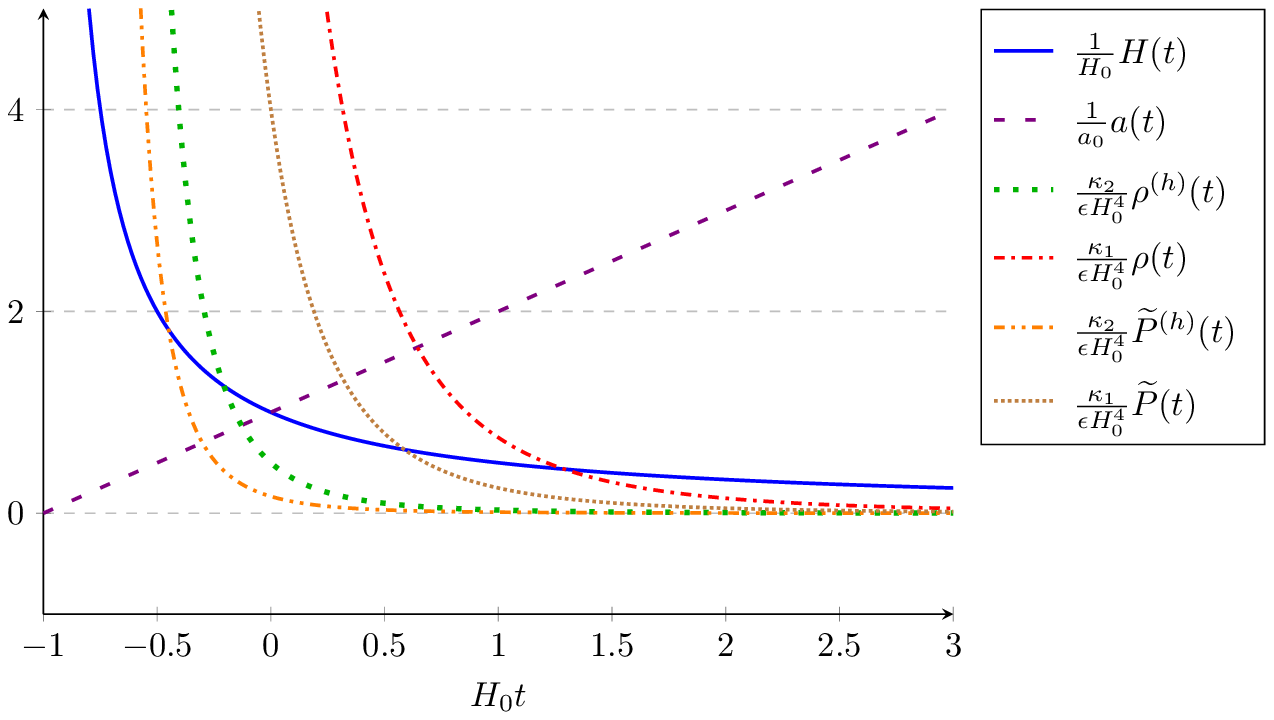}
\caption{Behavior of different cosmological parameters (vertical axis) in
terms of $t$, with $\widetilde{\omega}^{(h)}=1/3$, $n=1/2$, $\epsilon=1$ and $H_0=72$. Here we have $a=0$ at time $H_0t_b=-4/\left(3(\widetilde\omega^{(h)}+1)\right)=-1$.\label{fig07}}
\end{figure}

\begin{figure}[tbp]
%figure8
\centering
\includegraphics[width=0.7\columnwidth]{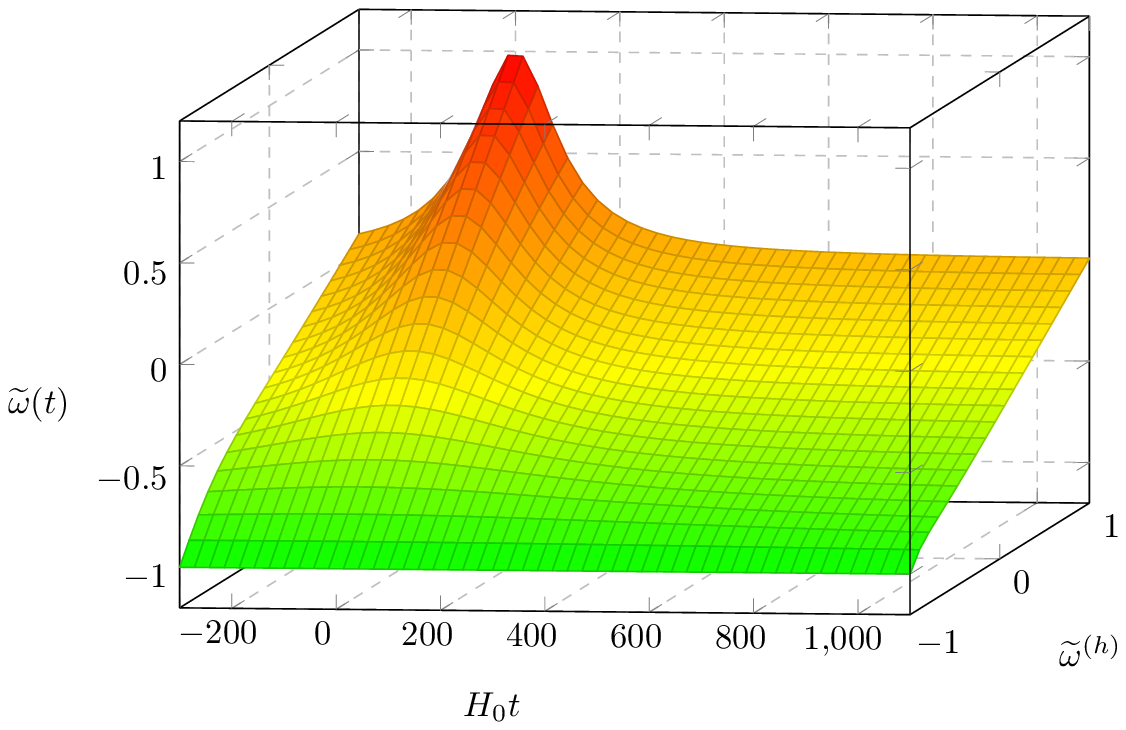}
\caption{Behavior of $\widetilde\omega(t)$ in terms of $t$
and $-1<\widetilde\omega^{(h)}<1$, with $n=1/2$, $\epsilon=1$ and $H_0=72$.\label{fig08}}
\end{figure}

\item For $n>1$ we have that in the case $\widetilde\omega^{(h)}<-1$, the equation (\ref{sf5d4}) has a similar behavior to the equation which describes an accelerated universe with a big rip-like future singularities at time 
\begin{equation*} %\label{tr01}
t_{r}=-\frac{4}{3(\widetilde{\omega }^{(h)}+1)H_{0}}.
\end{equation*}
It should note that the variable state parameter blows up in a finite future time. This behavior is similar to the w-singularities \cite{wrip1,wrip2}. However the matter energy density $\rho (t)$ is negative. For this reason we discard this solution. When $-1<\widetilde{\omega }^{(h)}<0$\ there is not big rip and the variable state parameter blows up after a while during its evolution. However the matter energy density $\rho (t)$ is negative. For this reason we discard this solution.

\end{enumerate}

Note that in the $\epsilon =0$ limit, we get the results of the standard cosmology for a scenario type big rip in the case of a fluid where: \begin{inparaenum}[(i)] 
\item the state parameter $\omega (t)=(\widetilde\omega ^{(h)}+1)/2$ is constant and
\item the scale factor (\ref{sf5d4}) is given as an ansatz. The difference lies in that in standard cosmology the exponent on the right side of the scale factor is $2/3(\omega +1)$ \cite{Copeland:2006wr}. Cosmology ECHS modifies this result: the exponent of the scale factor is $4/3(\widetilde\omega^{(h)}+1)$.
\end{inparaenum}

It might be interesting to mention that in this case we have obtained the exact form of the scale factor without having to make an expansion to the scale factor as was done in ref. \cite{gms}. This means that we have found a different solution that complements that found for $\omega \neq -1$ in ref. \cite{gms}.

\subsection{The $h^a$ fluid as a politropic fluid}

Now we consider the case where the field $h^{a}$ can be a polytropic fluid modelling dark energy. The corresponding state equation is given by 
\begin{equation*}
\widetilde{P}^{(h)}=\widetilde{\omega}^{(h)}\rho ^{(h)\gamma },
\end{equation*}%
where $\widetilde\omega ^{(h)}$ is the constant state parameter, $\gamma $ is the politropic index with $\gamma>0$ and $\gamma \neq 1$, because when $\gamma=1$, the equation of state is reduced to the barotropic equation of state \cite{poly1,poly2,poly3,poly4,poly5,poly6}. The matter field will be modeled as a fluid whose state equation with variable parameter, is given by equation (\ref{eosm5d4}).

We should note that when working with two fluids: is found that in
4-dimensional EChS cosmology, the cosmological variable are determined as a function of time, while in standard cosmology these variables are given in terms of the scale factor without being able to know its explicit form.

In this subsection we study two cases having exact solution: $\gamma=\frac {1}{2}$ and $\gamma=\frac{3}{4}$.

\subsubsection{Case 1: $\gamma=\frac{1}{2}$}

Writing equation (\ref{fe5da45}) in terms of the Hubble parameter we find 
\begin{equation}
\dot{H}+\frac{3}{4}H^{2}+\frac{3}{4}\,\widetilde{\omega}^{(h)}\left( \frac{\kappa
_{2}}{\epsilon n}\right) ^{\frac{1}{2}}=0  \label{poli4d1}
\end{equation}%
where we can see that it is not possible to consider the limit $\epsilon =0$. To find a solution for (\ref{poli4d1}), we have three different cases. First, when $\widetilde\omega ^{(h)}>0$ the scalar factor has a sector where it is negative, so that we discard this solution. Second, if $\widetilde\omega ^{(h)}=0$ then
\begin{equation*}
H(t)=\frac{H_{0}}{1+\frac{3}{4}H_{0}t},
\end{equation*}%
where it was considered $H_{0}=H(t=0)$. The scalar factor is given by 
\begin{equation*}
a(t)=a_{0}\left(1+\frac{3}{4}H_{0}t\right)^{\frac{4}{3}},
\end{equation*}%
where was considered $a_{0}=a(t=0)$. From equations (\ref{fe5da4}) and (\ref{fe5da41}) we find that $\rho^{(h)}$ and $\rho$ are given by
\begin{align*}
\rho^{(h)}(t)&=\frac{\epsilon nH_{0}^{4}}{\kappa_{2}\left(1+\frac{3}{4}H_{0}t\right)^{4}},\\
\rho(t)&=\frac{6H_{0}^{2}\left( (1-n)\bigl(\frac{3}{4}%
H_{0}t+1\bigr)^{2}+4\epsilon nH_{0}^{2}\right) }{\kappa_{1}\left(1+\frac{3}{4}H_{0}t\right)^{4}},
\end{align*}
so that the state parameter is then given by
\begin{equation*}
\widetilde{\omega}(t)=-\frac{(1-n)\left(1+\frac{3}{4}H_{0}t\right)^2}{2\left((1-n)\left(1+\frac{3}{4}H_{0}t\right)^2+4\epsilon n H_0^2\right)}
\end{equation*}

When $\widetilde\omega^{(h)}<0$, we find a solution given by 
\begin{equation*}
H(t)=H_4\tanh\left(\frac{3}{4}H_4\left(t-t_4\right) \right),
\end{equation*}
where
\begin{equation*}
H_4= \sqrt{-\widetilde\omega^{(h)}}\left(\frac{\kappa_2}{\epsilon n}\right)^{\frac{1}{4}},\quad t_4=-\frac{4}{3H_4}\text{arctanh}\left(\frac{H_0}{H_4}\right),
\end{equation*}
so that the scale factor is given by
\begin{equation*}
a(t)=a_0\frac{\cosh^{\frac{4}{3}}\left(\frac{3}{4}H_4\left(t-t_4\right) \right)}{\cosh^{\frac{4}{3}}\left(\frac{3}{4}H_4t_4 \right) } 
\end{equation*}
and from equations (\ref{fe5da4}) and (\ref{fe5da41}) we find that $\rho^{(h)}$ and $\rho$ are given by 
\begin{align*}
\rho^{(h)}(t)&=\frac{\left(\widetilde\omega^{(h)}\right)^2}{\epsilon n}\tanh^4\left(\frac{3}{4}H_4\left(t-t_4\right) \right) \\
\rho(t) & =\frac{-6\,\widetilde\omega^{(h)}\sqrt{\frac{\kappa_2}{\epsilon n}}}{\kappa_1}\tanh^2\left(\frac{3}{4}H_4\left(t-t_4\right) \right)\left(n-1+4\,\widetilde\omega^{(h)}\sqrt{\kappa_2\epsilon\,n}\tanh^2\left(\frac{3}{4}H_4\left(t-t_4\right) \right)\right).
\end{align*}

Therefore the state parameter is then given by
\begin{equation*}
\widetilde{\omega}(t)=\frac{\Bigl(\coth^2\bigr(\frac{3}{4}H_4\left(t-t_4\right) \bigl)+1\Bigr)\left(1-n\right)-8\widetilde\omega^{(h)}\sqrt{\kappa_2\epsilon\,n}}{2\Bigl(n-1+4\,\widetilde\omega^{(h)}\sqrt{\kappa_2\epsilon\,n}\,\tanh^2\left(\frac{3}{4}H_4\left(t-t_4\right) \right)\Bigr)}.
\end{equation*}

\subsubsection{Case 2: $\gamma=\frac{3}{4}$}

Writing equation (\ref{fe5da45}) in terms of the Hubble parameter we find 
\begin{equation*}
\dot{H}+\frac{3}{4}H^{2}+\frac{3}{4}\widetilde{\omega}^{(h)}\left( \frac{\kappa_{2}}{\epsilon n}\right)^{\frac{1}{4}}H=0
\end{equation*}where we can see that it is not possible to consider the limit $\epsilon=0$. The solution is given by 
\begin{equation*}
H(t)=\frac{H_5}{\exp\Bigl(\frac{3}{4}H_5(t-t_5)\Bigr)-1}
\end{equation*}%
where
\begin{equation*}
H_5=\widetilde{\omega}^{(h)}\left( \frac{\kappa_{2}}{\epsilon n}\right)^{\frac{1}{4}},\quad t_5=-\frac{4}{3H_5}\ln\left(1+\frac{H_5}{H_0}\right).
\end{equation*}

This means that the scale factor is
\begin{equation*}
a(t)=a_0\left(\frac{1-\exp\Bigl(-\frac{3}{4}H_5(t-t_5)\Bigr)}{1-\exp\Bigl(\frac{3}{4}H_5t_5\Bigr)}\right)^{\frac{4}{3}}
\end{equation*}
From (\ref{fe5da4}) and (\ref{fe5da41}) we can see that $\rho^{(h)}$ and $\rho$ are given by 
\begin{align*}
\rho^{(h)}(t)&=\left(\frac{\widetilde{\omega}^{(h)}}{\exp\Bigl(\frac{3}{4}H_5(t-t_5)\Bigr)-1}\right) ^{4}\\
\rho(t)&=\frac{6\left(\widetilde{\omega}^{(h)}\right)^2\sqrt{\frac{\kappa_2}{\epsilon n}} \left( (1-n)\left(\exp\Bigl(\frac{3}{4}H_5(t-t_5)\Bigr)-1\right)^{2}+4\sqrt{\kappa_{2}\epsilon n}\left(\widetilde{\omega}^{(h)}\right)^2\right)}{\kappa_1\left(\exp\Bigl(\frac{3}{4}H_5(t-t_5)\Bigr)-1\right)^{4}}
\end{align*}
Therefore the state parameter is then given by
\begin{equation*}
\widetilde{\omega}(t)=\frac{\left(e^{\frac{3}{4}H_5(t-t_5)}-1\right)\biggl(\left(e^{\frac{3}{4}H_5(t-t_5)}-1\right)\left(e^{\frac{3}{4}H_5(t-t_5)}-2\right)(1-n)+8\sqrt{\kappa_2\epsilon n}\left(\widetilde{\omega}^{(h)}\right)^2\biggr)}{2\left(\left(e^{\frac{3}{4}H_5(t-t_5)}-1\right)^2(1-n)+4\sqrt{\kappa_2\epsilon n}\left(\widetilde{\omega}^{(h)}\right)^2\right)}.
\end{equation*}

\subsection{The $h^a$ fluid as a variable modified Chaplygin gas}

Now consider the case where the field $h^{a}$ can be a variable modified Chaplygin gas modelling dark energy. The corresponding state equation is given by 
\begin{equation*}
\widetilde{P}^{(h)}=-\frac{a(t)^{-\beta }}{\rho ^{(h)\gamma }},
\end{equation*}%
where $0<\gamma <1$ and $\beta >0$ \cite{chap1,chap2,chap3,chap4,chap5,chap6,chap7,chap8,chap9,chapkk}. The matter field will be modeled as a fluid whose state equation with variable parameter, is given by equation (\ref{eosm5d4}). Using this state equation it is direct to see that 
\begin{equation*}
\dot{z}+3\frac{\dot{a}}{a}(\gamma +1)z=3(\gamma +1)\frac{\dot{a}}{a^{\beta
+1}}
\end{equation*}%
where $z=\rho ^{(h)\gamma +1}$. So that 
\begin{equation*}
\rho ^{(h)}(t)=\left(\frac{3(\gamma +1)}{3(\gamma +1)-\beta }a(t)^{-\beta
}+a_{6}a(t)^{-3(\gamma +1)}\right) ^{\frac{1}{\gamma +1}}.
\end{equation*}%
From this equation and (\ref{fe5da4}) we find 
\begin{equation}
H(t)=H_6\left(\frac{3(\gamma +1)}{3(\gamma
+1)-\beta }a(t)^{-\beta }+a_{6}a(t)^{-3(\gamma +1)}\right) ^{\frac{1}{4(\gamma +1)}}  \label{ten}
\end{equation}
where $H_6$ and the constant of integration $a_6$ are given by
\begin{equation*}
H_6=\left(\frac{\kappa _{2}}{n\epsilon }\right)^{\frac{1}{4}},\quad
a_6=a_0^{3(\gamma+1)}\left(\left(\frac{H_0}{H_6}\right)^{\gamma+1}-\frac{3(\gamma+1)}{3(\gamma+1)-\beta}a_0^{-\beta}\right).
\end{equation*}

Introducing (\ref{ten}) in (\ref{fe5da41}) we can see that
\begin{align*}
\rho (t)=&\frac{6\sqrt{\frac{\kappa_2}{\epsilon n}}}{\kappa_1}\left(\frac{3(\gamma +1)}{3(\gamma
+1)-\beta }a(t)^{-\beta }+a_{6}a(t)^{-3(\gamma +1)}\right)^{\frac{1}{2(\gamma +1)}}\notag \\
&\times\left(1-n+4\sqrt{\kappa_2\epsilon n}\left(\frac{3(\gamma +1)}{3(\gamma
+1)-\beta }a(t)^{-\beta }+a_{6}a(t)^{-3(\gamma +1)}\right) ^{\frac{1}{2(\gamma +1)}} \right)
\end{align*}

We should note that if $a_6=0$ in equation (\ref{ten}) we find a particular solution for the scale factor
\begin{equation}
a(t)=\left(\frac{3(\gamma+1)}{3(\gamma+1)-\beta}\right)^{\frac{1}{\beta}}\left(\frac{4(\gamma+1)}{\beta}H_6(t-t_6)\right)^{\frac{4(\gamma+1)}{\beta}},  \label{85}
\end{equation}
where 
\begin{equation*}
t_6=-\frac{\beta}{4H_6(\gamma+1)}\left(\frac{3(\gamma+1)}{3(\gamma+1)-\beta}\right)^{-\frac{1}{4(\gamma+1)}}a_0^{\frac{\beta}{4(\gamma+1)}}
\end{equation*}

From equation (\ref{85}) we can see that if $\frac{4(\gamma+1)}{\beta }>1$, that is $0<\beta <4$ and $0<\gamma <1$, or $4<\beta <8$ and $\frac{1}{4}(\beta -4)<\gamma <1$, then we have an accelerated power law type solution. Therefore the state parameter is then given by
\begin{equation*}
\widetilde{\omega}(t)=\frac{(1-n)  \left(\frac{4(\gamma+1)}{\beta}H_6(t-t_6)\right)^{\frac{8(\gamma+1)}{\beta}}+\frac{2048}{\beta^4}\sqrt{\kappa_2\epsilon n}(\gamma+1)^4}{\frac{6(\gamma+1)}{\beta}\left((1-n)\left(\frac{4(\gamma+1)}{\beta}H_6(t-t_6)\right)^{\frac{8(\gamma+1)}{\beta}}+\frac{1024}{\beta^4}\sqrt{\kappa_2\epsilon n}(\gamma+1)^4\right)}-1.
\end{equation*}

\section{Concluding Remarks}

We have found some new cosmological solutions for the so called Einstein-Chern-Simons-Friedmann-Robertson-Walker field equations taking the $h^a$ field component as modelling the dark energy component, represented by three different types of equations of state: barotropic, polytropic, and varying generalized modified Chaplygin gas.

For the three types of equations of state, it was found that the behavior of matter field is described by a barotropic equation of state, with variable state parameter $P=\omega (t)\rho$.

A comment about the differences and similarities of the results found in this work and the known results of standard cosmology could be of interest.

\begin{enumerate} [(i)]
\item If the behavior of two fluids is studied and if we want describe, in the context of standard cosmology, a universe accelerated at a late stage, then one of the fluids may represent dark matter described by the equation of state $P_{1}=\omega_{1}\rho _{1}$ (con $\omega_{1}=0$), while the other may represent dark energy described by the equation of state $P_{2}=\omega_{2}\rho_{2}$ (con $\omega_{2}<-1/3$).

In this case we can consider that each fluid evolves independently of each other or take the option of considering that there is an interaction term between the two fluids $Q$, this means that fluids will not evolve independently and behavior of a fluid depend on the behavior of the other.

\item If in the context of cosmology Einstein-Chern-Simons we consider that the matter field of density $\rho$ represents dark matter (or dark energy) and that the $h^a$ field of density $\rho^{(h)}$ represents dark energy (or dark matter) then, if the two fluids evolve independently of each other, we find that the behavior of a fluid depends on the behavior of the other fluid through the geometric term $H^{4}$. This term plays the role of the interaction $Q$ term present in the case of standard cosmology.

\item If in standard cosmology we consider two fluids, one with a constant state parameter and the other with a variable state parameter then, to solve the system dynamics it is necessary to specify the state equations for each fluid and consider an ansatz for the variable parameter state, giving the scale factor or giving an energy density.

\item In Einstein-Chern-Simons cosmology, this situation can be resolved without considering an ansatz, closing down the system and getting the form of the different cosmological variables.
\end{enumerate}

\begin{acknowledgments}
This work was supported in part by FONDECYT Grants No.$1130653$ and No.$1150719$ from the Government of Chile. Two of the authors were supported by FONDECYT Grants No.$3130444$ (P. M.) and Comisi\'on Nacional de Investigaci\'{o}n Cient\'{\i}fica y Tecnol\'{o}gica CONICYT  No.$21160827$ (L. A.) and from Universidad de Concepci\'{o}n, Chile. C. Q. was supported by Direcci\'on de Investigaci\'on de la Universidad San Sebasti\'an through Grant VRA-USS 2015-0018-I.
\end{acknowledgments}


\begin{thebibliography}{99}    
\bibitem{gms}
F.~Gomez, P.~Minning and P.~Salgado, \emph{Standard cosmology in Chern-Simons gravity}, \href{http://dx.doi.org/10.1103/PhysRevD.84.063506}{\emph{Phys. Rev. D} {\bf 84} (2011) 063506}.
  
\bibitem{cgqs}
M.~Cataldo, J.~Cris{\'{o}}stomo, S.~del Campo, F.~G{\'{o}}mez, C.~C. Quinzacara
  and P.~Salgado, \emph{Accelerated FRW solutions in Chern-Simons gravity},
  \href{http://dx.doi.org/10.1140/epjc/s10052-014-3087-9}{\emph{Eur. Phys. J. C} {\bf 74} (2014) 3087},
  [\href{http://arxiv.org/abs/1401.2128}{{\tt arXiv:1401.2128}}].
  
\bibitem{champ1}
A.~Chamseddine, \emph{Topological gauge theory of gravity in five and all odd
  dimensions},
  \href{http://dx.doi.org/10.1016/0370-2693(89)91312-9}{\emph{Phys. Lett. B} {\bf 233} (1989) 291--294}.

\bibitem{champ2}
A.~Chamseddine, \emph{Topological gravity and supergravity in various
  dimensions},
  \href{http://dx.doi.org/10.1016/0550-3213(90)90245-9}{\emph{Nucl. Phys.
  B} {\bf 346} (1990) 213--234}.

\bibitem{zan1}
J.~Zanelli, \emph{{Lecture notes on Chern-Simons (super-)gravities. Second
  edition (February 2008)}},  [\href{http://arxiv.org/hep-th/0502193}{{\tt
  arXiv:hep-th/0502193}}].

\bibitem{salg1}
F.~Izaurieta, P.~Minning, A.~Perez, E.~Rodriguez and P.~Salgado,
  \emph{{Standard general relativity from Chern-Simons gravity}},
  \href{http://dx.doi.org/10.1016/j.physletb.2009.06.017}{\emph{Phys.
  Lett. B}
  {\bf 678} (2009) 213--217}, [\href{http://arxiv.org/abs/0905.2187}{{\tt
  arXiv:0905.2187}}].

\bibitem{salg2}
F.~Izaurieta, E.~Rodr{\'{i}}guez, P.~Salgado,
  \emph{{Expanding Lie (super)algebras through abelian semigroups}},
  \href{http://dx.doi.org/10.1063/1.2390659}{\emph{J. Math. Phys.} {\bf 47} (2006) 42}, [\href{http://arxiv.org/abs/hep-th/0606215}{{\tt
  arXiv:hep-th/0606215}}].

\bibitem{salg3}
F.~Izaurieta, A.~Perez, E.~Rodriguez and P.~Salgado, \emph{{Dual formulation of the Lie algebra S-expansion procedure}},
  \href{http://dx.doi.org/10.1063/1.3171923}{\emph{J. Math. Phys.} {\bf 50} (2009) 073511},
  [\href{http://arxiv.org/abs/0903.4712}{{\tt arXiv:0903.4712}}].

\bibitem{azcarr}
J.~de~Azc{\'{a}}rraga, J.~Izquierdo, M.~Pic{\'{o}}n and O.~Varela,
  \emph{{Generating Lie and gauge free differential (super)algebras by
  expanding Maurer-Cartan forms and Chern-Simons supergravity}},
  \href{http://dx.doi.org/10.1016/S0550-3213(03)00342-0}{\emph{Nucl. Phys.
  B} {\bf 662} (2003) 185--219}, [\href{https://arxiv.org/abs/hep-th/0212347}{{\tt arXiv:hep-th/0212347}}].

\bibitem{poly1}
K.~Karami, S.~Ghaffari and J.~Fehri, \emph{{Interacting polytropic gas model of phantom dark energy in non-flat universe}},
  \href{http://dx.doi.org/10.1140/epjc/s10052-009-1120-1}{\emph{Eur.
  Phys. J. C} {\bf 64} (2009) 85--88},
  [\href{http://arxiv.org/abs/0911.4915}{{\tt arXiv:0911.4915}}].

\bibitem{poly2}
M.~Malekjani and A.~Khodam-Mohammadi, \emph{{Statefinder Diagnostic and
  $w-w'$ Analysis for Interacting Polytropic Gas Dark Energy Model}},
  \href{http://dx.doi.org/10.1007/s10773-012-1195-6}{\emph{Int.
  J. Theor. Phys.} {\bf 51} (2012) 3141--3151},
  [\href{http://arxiv.org/abs/1201.0589}{{\tt arXiv:1201.0589}}].

\bibitem{poly3}
M.~Malekjani, \emph{{Polytropic Gas Scalar Field Models of Dark Energy}},
  \href{http://dx.doi.org/10.1007/s10773-013-1558-7}{\emph{Int.
  J. Theor. Phys.} {\bf 52} (2013) 2674--2685},
  [\href{http://arxiv.org/abs/1206.0647}{{\tt arXiv:1206.0647}}].

\bibitem{poly4}
M.~Taji and M.~Malekjani, \emph{{Interacting Holographic Polytropic Gas Model of Dark Energy}},
  \href{http://dx.doi.org/10.1007/s10773-013-1641-0}{\emph{Int.
  J. Theor. Phys.} {\bf 52} (2013) 3405--3412}.

\bibitem{poly5}
S.~Asadzadeh, Z.~Safari, K.~Karami and A.~Abdolmaleki, \emph{{Cosmological
  Constraints on Polytropic Gas Model}},
  \href{http://dx.doi.org/10.1007/s10773-013-1922-7}{\emph{Int.
  J. Theor. Phys.} {\bf 53} (2014) 1248--1262},
  [\href{http://arxiv.org/abs/1209.6374}{{\tt arXiv:1209.6374}}].

\bibitem{poly6}
K.~Kleidis and N.~K. Spyrou, \emph{{Polytropic dark matter flows illuminate
  dark energy and accelerated expansion}},
  \href{http://dx.doi.org/10.1051/0004-6361/201424402}{\emph{Astron.
  Astrophys.} {\bf 576} (2015) A23},
  [\href{http://arxiv.org/abs/1411.6789}{{\tt arXiv:1411.6789}}].

\bibitem{chap1}
Z.-K. Guo and Y.-Z. Zhang, \emph{{Cosmology with a variable Chaplygin gas}},
  \href{http://dx.doi.org/10.1016/j.physletb.2006.12.063}{\emph{Phys. Lett.
  B} {\bf 645} (2007) 326--329}, [\href{https://arxiv.org/abs/astro-ph/0506091}{{\tt
  arXiv:astro-ph/0506091}}].

\bibitem{chap2}
U.~Debnath, \emph{{Variable modified Chaplygin gas and accelerating universe}},
  \href{http://dx.doi.org/10.1007/s10509-007-9690-6}{\emph{Astrophys.
  Space Sci.} {\bf 312} (2007) 295--299},
  [\href{http://arxiv.org/abs/0710.1708}{{\tt arXiv:0710.1708}}].

\bibitem{chap3}
M.~Jamil and M.~A. Rashid, \emph{{Interacting modified variable Chaplygin gas in a non-flat universe}},
  \href{http://dx.doi.org/10.1140/epjc/s10052-008-0722-3}{\emph{Eur.
  Phys. J. C} {\bf 58} (2008) 111--114},
  [\href{http://arxiv.org/abs/0802.1146}{{\tt arXiv:0802.1146}}].

\bibitem{chap4}
L.~Xing, Y.~Gui, L.~Xu and J.~Lu, \emph{{Evolution of variable modified
  Chaplygin gas model}},
  \href{http://dx.doi.org/10.1142/S0217732309027443}{\emph{Mod. Phys.
  Lett. A} {\bf 24} (2009) 683--691}.

\bibitem{chap5}
S.~Chattopadhyay and U.~Debnath, \emph{{Holographic dark energy scenario and variable modified Chaplygin gas}},
  \href{http://dx.doi.org/10.1007/s10509-009-9977-x}{\emph{Astrophys. Space Sci.} {\bf 319} (2009) 183--185},
  [\href{http://arxiv.org/abs/0901.2184}{{\tt arXiv:0901.2184}}].

\bibitem{chap6}
M.~U. Farooq, M.~Jamil and M.~A. Rashid, \emph{{Interacting Entropy-Corrected Holographic Chaplygin Gas Model}},
  \href{http://dx.doi.org/10.1007/s10773-010-0420-4}{\emph{Int.
  J. Theor. Phys.} {\bf 49} (2010) 2334--2347},
  [\href{http://arxiv.org/abs/1003.3399}{{\tt arXiv:1003.3399}}].

\bibitem{chap7}
K.~Karami and A.~Abdolmaleki, \emph{{Polytropic and Chaplygin $f(T)$-gravity
  models}},
  \href{http://dx.doi.org/10.1088/1742-6596/375/1/032009}{\emph{J.
  Phys. Conf. Ser.} {\bf 375} (2012) 032009}.

\bibitem{chap8}
K.~Karami and M.~S. Khaledian, \emph{{Polytropic and Chaplygin $f(R)$-gravity
  models}},
  \href{http://dx.doi.org/10.1142/S0218271812500836}{\emph{Int.
  J. Mod. Phys. D} {\bf 21} (2010) 1250083},
  [\href{http://arxiv.org/abs/1010.2639}{{\tt arXiv:1010.2639}}].

\bibitem{chap9}
J.~Bhadra and U.~Debnath, \emph{{Dynamical system analysis of interacting
  variable modified Chaplygin gas model in FRW universe}},
  \href{http://dx.doi.org/10.1140/epjp/i2012-12030-2}{\emph{Eur.
  Phys. J. Plus} {\bf 127} (2012) 30},
  [\href{http://arxiv.org/abs/1109.3578}{{\tt arXiv:1109.3578}}].

\bibitem{chapkk}
C.~Ranjit, S.~Chakraborty and U.~Debnath, \emph{{Variable Modified Chaplygin Gas in Ani\-so\-tro\-pic Universe with Kaluza-Klein Metric}},
  \href{http://dx.doi.org/10.1007/s10773-012-1395-0}{\emph{Int.
  J. Theor. Phys.} {\bf 52} (2013) 862--876},
  [\href{http://arxiv.org/abs/1201.0852}{{\tt arXiv:1201.0852}}].

\bibitem{andrew}
K.~Andrew, B.~Bolen and C.~A. Middleton, \emph{{Solutions of higher dimensional Gauss-Bonnet FRW cosmology}},
  \href{http://dx.doi.org/10.1007/s10714-007-0502-7}{\emph{Gen. Relativ.
  Gravit.} {\bf 39} (2007) 2061--2071},
  [\href{http://arxiv.org/abs/0708.0373}{{\tt arXiv:0708.0373}}].

\bibitem{mohamedi}
N.~Mohammedi, \emph{{Dynamical compactification, standard cosmology, and the
  accelerating universe}},
  \href{http://dx.doi.org/10.1103/PhysRevD.65.104018}{\emph{Phys. Rev. D}
  {\bf 65} (2002) 104018}, [\href{https://arxiv.org/abs/hep-th/0202119}{{\tt
  arXiv:hep-th/0202119}}].

\bibitem{hub1}
W.~L. Freedman, B.~F. Madore, B.~K. Gibson, L.~Ferrarese, D.~D. Kelson,
  S.~Sakai et~al., \emph{{Final Results from the Hubble Space Telescope Key
  Project to Measure the Hubble Constant}},
  \href{http://dx.doi.org/10.1086/320638}{\emph{Astrophys. J.} {\bf
  553} (2001) 47--72}, [\href{http://arxiv.org/abs/astro-ph/0012376}{{\tt
  arXiv:astro-ph/0012376}}].

\bibitem{hub2}
G.~Hinshaw, J.~L. Weiland, R.~S. Hill, N.~Odegard, D.~Larson, C.~L. Bennett
  et~al., \emph{{Five-Year Wilkinson Microwave Anisotropy Probe (WMAP)
  Observations: Data Processing, Sky Maps, and Basic Results}},
  \href{http://dx.doi.org/10.1088/0067-0049/180/2/225}{\emph{Astrophys. J. Suppl. Ser.} {\bf 180} (2008) 225--245},
  [\href{http://arxiv.org/abs/0803.0732}{{\tt arXiv:0803.0732}}].
  
\bibitem{hub3}
N.~Jarosik, C.~L. Bennett, J.~Dunkley, B.~Gold, M.~R. Greason, M.~Halpern
  et~al., \emph{{Seven-Year Wilkinson Microwave Anisotropy Probe (WMAP)
  Observations: Sky Maps, Systematic Errors, and Basic Results}},
  \href{http://dx.doi.org/10.1088/0067-0049/192/2/14}{\emph{Astrophys. J. Suppl. Ser.} {\bf 192} (2010) 14},
  [\href{http://arxiv.org/abs/1001.4744}{{\tt arXiv:1001.4744}}].

\bibitem{hub4}
G.~Hinshaw, D.~Larson, E.~Komatsu, D.~N. Spergel, C.~L. Bennett, J.~Dunkley
  et~al., \emph{{Nine-Year Wilkinson Microwave Anisotropy Probe (WMAP)
  Observations: Cosmological Parameter Results}},
  \href{http://dx.doi.org/10.1088/0067-0049/208/2/19}{\emph{Astrophys. J. Suppl. Ser.} {\bf 208} (2012) 19},
  [\href{http://arxiv.org/abs/1212.5226}{{\tt arXiv:1212.5226}}].

\bibitem{wrip1}
M.~P. D{\c{a}}browski and T.~Denkiewicz, \emph{{Barotropic index
  $w$-singularities in cosmology}},
  \href{http://dx.doi.org/10.1103/PhysRevD.79.063521}{\emph{Phys. Rev. D}
  {\bf 79} (2009) 063521}, [\href{http://arxiv.org/abs/0902.3107}{{\tt
  arXiv:0902.3107}}].

\bibitem{wrip2}
L.~Fern{\'{a}}ndez-Jambrina, \emph{$w$-cosmological singularities},
  \href{http://dx.doi.org/10.1103/PhysRevD.82.124004}{\emph{Phys. Rev. D}
  {\bf 82} (2010) 124004}, [\href{http://arxiv.org/abs/1011.3656}{{\tt
  arXiv:1011.3656}}].

\bibitem{Copeland:2006wr}
E.~J. Copeland, M.~Sami and S.~Tsujikawa, \emph{{Dynamics of Dark Energy}},
  \href{http://dx.doi.org/10.1142/S021827180600942X}{\emph{Int.
  J. Mod. Phys. D} {\bf 15} (2006) 1753--1935},
  [\href{https://arxiv.org/abs/hep-th/0603057}{{\tt arXiv:hep-th/0603057}}].

\end{thebibliography}
\end{document}